\documentclass[letterpaper,prl,twocolumn,showpacs,preprintnumbers]{revtex4}

\usepackage{graphicx}
\usepackage{bm}


\begin{document}

\title{Spiral Exchange Coupling in Trilayer Magnetic Junction Mediated by Diluted-Magnetic-Semiconductor Thin Film }

\author{Chia-Hui Lin}
\affiliation{Department of Physics, National Tsing-Hua
University,Hsinchu 300, Taiwan}
\affiliation{Physics Division,
National Center for Theoretical Sciences, Hsinchu 300, Taiwan}

\author{Hsiu-Hau Lin}
\affiliation{Department of Physics, National Tsing-Hua
University,Hsinchu 300, Taiwan}
\affiliation{Physics Division,
National Center for Theoretical Sciences, Hsinchu 300, Taiwan}

\author{Tzay-Ming Hong}
\affiliation{Department of Physics, National Tsing-Hua
University,Hsinchu 300, Taiwan}
\affiliation{Physics Division,
National Center for Theoretical Sciences, Hsinchu 300, Taiwan}

\date{\today}

\begin{abstract}
We revisit the non-collinear exchange coupling across the trilayer magnetic junction mediated by the diluted-magnetic-semiconductor thin film. By numerical approaches, we investigate the spiral angle between the ferromagnetic layers extensively in the parameter space. In contrast to previous study, we discovered the important role of spin relaxation, which tends to favor spiral exchange over the oscillatory Ruderman-Kittel-Kasuya-Yosida interaction. Finally, we discuss the physics origins of these two types of magnetic interactions.
\end{abstract}
\maketitle

The goal to merge the functionalities of information processing and data storage in one single material has charmed researchers in the field of spintronics for many years.\cite{Wolf01,MacDonald05} One of the promising candidates is the diluted magnetic semiconductor (DMS), made of the III-V host semiconductor doped with transition metals, such as (Ga,Mn)As. The ferromagnetic order in DMS\cite{Akai98,Dietl00,Konig00,Schliemann01,Litvinov01} arises from the aligned magnetic moments of the transition metals, mediated by itinerant carriers in the semiconducting bands and thus can be easily manipulated by electrical means. This unique feature has attracted enormous interests both in academic research and potential industrial applications.

\begin{figure}
\centering
\includegraphics[width=6cm]{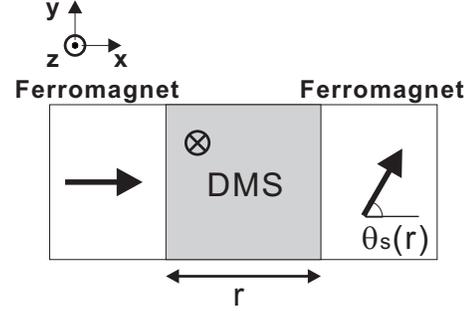}
\caption{\label{fig1} Schematic figure for the F/DMS/F trilayer magnetic junction, where the spiral angle $\theta_{s}(r)$ denotes the relative angle between the two ferromagnets.}
\end{figure}

In addition to its potential applications for making the next-generation transistors in spintronics, DMS also brings up surprises in the more conventional magnetic multilayers as shown in Fig.~\ref{fig1}. One of the authors studied the F/DMS/F trilayer magnetic junction, within the linear response theory\cite{Sun04} and the self-consistent Green's function approach\cite{Yang01,Konig01,Sun04a}, and found interesting spiral exchange coupling between the ferromagnets. Their numerical results show that the spiral exchange always beats the Ruderman-Kittel-Kasuya-Yosida (RKKY) interaction, rendering it into decorative ripples on the spiral backbones. Being puzzled by the dominance of the spiral exchange, we revisited the problem and found the important ingredient overlooked in the previous study -- the proper inclusion of the spin relaxation rate.

In this Letter, employing the same model described in Ref.\cite{Sun04}, we studied the trilayer magnetic junction within linear response theory. We carried out extensive numerical computations in different parameter regimes, in particular, with different spin relaxation rates $\eta$ (which was fixed in the previous study). We found that the competition between the spiral exchange and the RKKY interaction sensitively depends on the strength of spin relaxation. In the extreme ballistic limit, the RKKY interaction prevails, while the spiral exchange starts to take over the leading role when approaching the diffusive regime. The value of $\eta$ used in Ref.\cite{Sun04} happened to set the system in the diffusive regime (see below) where the spiral exchange dominates. In general, depending on the spin relaxation rate, it is possible to observe {\em both} types of magnetic behaviors!

Before diving into numerical details, a simple argument from the single-particle picture would help readers to understand the origin of the spiral exchange. In fact, the argument closely parallels Datta-Das' original proposal\cite{Datta90} for a spin field-effect transistor. Imagine an electron with $+x$ spin orientation is injected into the DMS thin layer from the ferromagnet on the left-hand side. The two components of the wave function evolve differently because of the Zeeman splitting in DMS,
\begin{eqnarray}
\Psi(0) = \frac{1}{\sqrt{2}}\left(\begin{array}{c}
1\\
1
\end{array}\right) \to 
\Psi(r) = \frac{1}{\sqrt{2}}\left(\begin{array}{c}
e^{ik_{F\uparrow}r}\\
e^{ik_{F\downarrow}r}
\end{array}\right),
\end{eqnarray}
here $r$ is the thickness of the DMS thin layer. The phase difference between the spinor components indicates that the carrier-mediated exchange coupling is non-collinear and tends to align the other ferromagnet on the right at a different angle,
$\theta_s(r) = - 2\pi r/\lambda_{s} = -(k_{F\uparrow} - k_{F\downarrow})r$. However, this simple single-particle picture may not be the whole story because it does not capture the quantum interferences from different patches of the whole Fermi surface, which give rise to the oscillatory RKKY interaction. Therefore, to pin down the dominant magnetic interaction, one needs to resort to the more formal Green's function approach.

After integrating out the itinerant carriers, it can be shown that the effective exchange coupling between the ferromagnets is proportional to the static spin susceptibility,
\begin{eqnarray}
\chi^{ij} (r) = \int^{\infty}_{0} e^{-\eta t} \langle\langle i [\sigma^i (\vec{r},t),\sigma^j(0,0)] \rangle\rangle dt ,
\label{chi}
\end{eqnarray}
where $\sigma^i$ is the spin density in DMS layer and the double bracket denotes thermal and quantum mechanical averages. Note that the vector dependence can be dropped because of the external SO(3) symmetry for the spatial coordinates (not to be confused with the internal SO(2) symmetry for the spinor). Furthermore, a phenomenology parameter $\eta$ is introduced to describe the spin relaxation in DMS layer.

In the presence of the finite Zeeman gap, the spinor has an internal SO(2) symmetry. Consider a $\pi/2$-rotation along the $z$-axis (always chosen to be the quantization axis for the Zeeman splitting). It changes $(S_{x},S_{y}) \to (S_{y},-S_{x})$ and leaves $S_{z}$ intact, implying $\chi^{xy}= -\chi^{yx}$ and $\chi^{xx}=\chi^{yy}$. Similar argument leads to $\chi^{xz}=0=\chi^{yz}$. For general geometries, the orientation of the pinned ferromagnet can be $\bm{n}_L = (\sin\theta\cos\phi, \sin\theta\sin\phi, \cos\theta)$.  Mediated by the itinerant carriers, it will lock the free magnetic moment on the right-hand side in the direction of $\bm{n}_R$,
\begin{eqnarray}\label{eq:angle}
\left[\begin{array}{c}
n_R^x\\
n_R^y\\
n_R^z
\end{array} \right] = \left[ \begin{array}{ccc}
\chi^{xx}&\chi^{xy}&0\\
-\chi^{xy}&\chi^{xx}&0\\
0&0&\chi^{zz}
\end{array}\right]
\left[\begin{array}{c}
n_L^x\\
n_L^y\\
n_L^z
\end{array} \right].
\end{eqnarray}
Since our focus here is to determine the dominance of different magnetic interactions, we choose the simplest geometry in Fig.~\ref{fig1} by setting $\bm{n}_L=(1,0,0)$. Thus, it is clear that $\bm{n}_R$ can be described by the spiral angle $\theta_s(r) =-\tan^{-1}[\chi^{xy}(r)/\chi^{xx}(r)]$. To find out the spiral angle,  we use the well-known trick to compute the complex susceptibility $\chi^{+-} = \chi^{xx}+i\chi^{xy}$, then extract the desired real and imaginary parts,
\begin{eqnarray}
\chi^{+-}(r) = \sum_{k k'}
\frac{f_{\uparrow}(\epsilon_{k}) - f_{\downarrow}(\epsilon_{k'})}{\epsilon_{k}-\epsilon_{k'}-\Delta+i\eta}
\, e^{ i( \vec{k} -\vec{k'}) \cdot\vec{r}},
\label{complexchi}
\end{eqnarray}
where $\epsilon_k = k^2/2m^*$ is the dispersion for itinerant carriers and $\Delta$ is the Zeeman gap. The Fermi-Dirac functions for the itinerant carriers are $f_{\alpha}(\epsilon_{k}) = [\exp(\epsilon_{k}-\alpha \Delta-\mu)+1]^{-1}$, with $\alpha = \pm$ corresponding to up/down spin flavors.

When carrying out numerical calculations for the spin susceptibility, we found that $\chi_{xy} \sim \eta^2$ vanishes as $\eta \to 0^+$. It implies that RKKY dominates in the ideal ballistic limit. Actually, in this limit, the spin-spin commutator in Eq.~(\ref{chi}) can be computed analytically and is indeed zero. We later realized that this result is due to a less obvious time-reversal symmetry within each spin flavor. It gives rise to the modified Onsager relation $\chi^{xy}=\chi^{yx}$ and forces the off-diagonal component $\chi_{xy}=0$ in the ideal ballistic limit. In Ref.\cite{Sun04}, the authors only chose one particular $\eta$ for their numerical computations (which happens to be in the diffusive regime). Therefore, their conclusions are incorrect in the extreme ballistic limit.

\begin{figure}
\centering
\includegraphics[width=6cm]{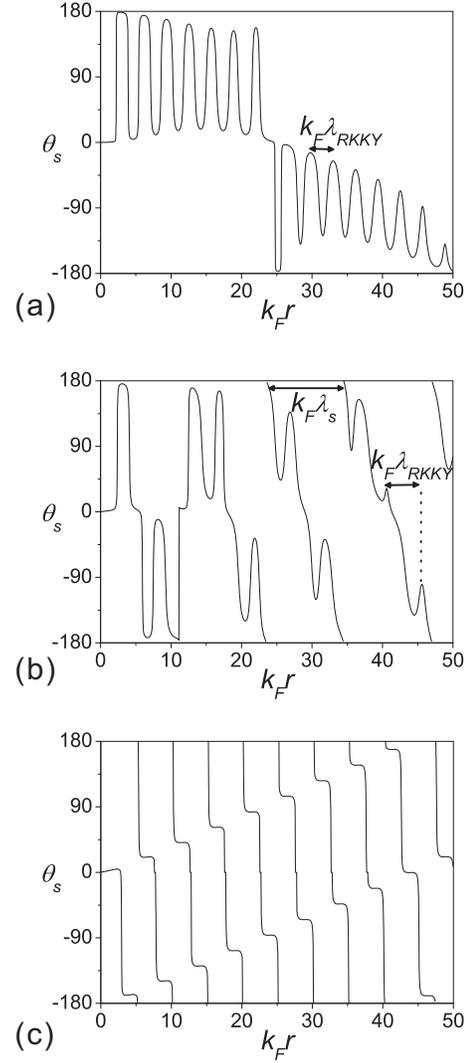}
\caption{Spiral angle $\theta_s(r)$ for (a) $k_{\Delta}/k_F=0.5$ (b) $k_{\Delta}/k_F=1.0$ (c) $k_{\Delta}/k_F=1.3$ in the ballistic regime with $k_{\eta}/k_F=0.23$.}
\label{fig2}
\end{figure}

\begin{figure}
    \centering
   \includegraphics[width=6cm]{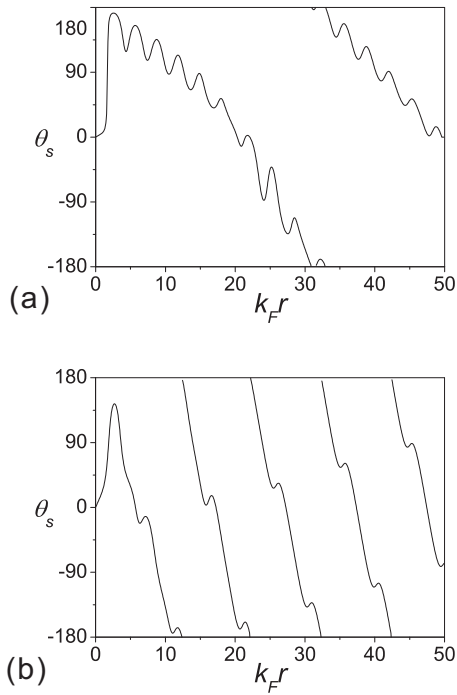}
  \caption{$\theta_s(r)$ for $k_{\eta}/k_F=1.0$ in the diffusion regime:
  (a) $(k_{\Delta}/k_F)=0.5$,
  (b) $(k_{\Delta}/k_F)=1.0$.}
  \label{fig3}
\end{figure}

In realistic materials, the spin relaxation rate $\eta$ is finite and the symmetry constraint no longer applies. By gradually increasing $\eta$ from the ballistic to the diffusive regimes, we show how the trend of the spiral angle $\theta_{s}(r)$ changes with it, as summarized in Figs. \ref{fig2} and \ref{fig3}. For convenience, we introduce $k_{\Delta} \equiv \sqrt{2m^{*} \Delta}$ to denote the inverse-length scale for the Zeeman gap and $k_{\eta} \equiv \sqrt{2m^{*} \eta}$ for spin relaxation. The total density of the itinerant carriers is also converted to $k_{F}$. In DMS, $k_{\Delta}$ has a sensitive dependence on the temperature through the Zeeman gap and can be as large as (or larger than) $k_{F}$ (roughly 1/nm in typical DMS materials) at low temperatures. Furthermore, the spin lifetime is about $10^{-2}$ picosecond, giving the ratio $k_{\eta}/k_F \approx 0.23$.

Making use of this rough estimate, we compute the spiral angle with different Zeeman splitting $k_{\Delta}$, shown in Fig. \ref{fig2}. When the polarization is small ($k_{\Delta}/k_{F} =0.5$), the signature of RKKY oscillations is rather obvious. On the other hand, when the carriers are fully polarized ($k_{\Delta}/k_{F} =1.3$), the spiral rotation becomes transparent and the RKKY oscillations, although visible, are suppressed into minor ripples. Careful analysis shows that these complicated patterns for the spiral angle can be characterized by two length scales,
\begin{eqnarray}
\lambda_{RKKY} = \frac{2\pi}{k_{F\uparrow}+k_{F\downarrow}},
\quad
\lambda_{s}    = \frac{2\pi}{k_{F\uparrow}-k_{F\downarrow}}.
\label{eq:lamda:spiral}
\end{eqnarray}
These length scales originate from the low-energy spin excitations. Consider the particle-hole excitation by kicking a spin-down electron to the spin-up band. It will carry momentum $\vec{p} = \vec{k}_{F\uparrow}-\vec{k}_{F\downarrow}$. The length scale $\lambda_s$ arises from excitations with $k_{F\uparrow}$ and $k_{F\downarrow}$ parallel, while $\lambda_{RKKY}$ comes from those with antiparallel momenta. The competition between these two type of spin excitations leads to the complicated patterns of the spiral angle $\theta_{s}(r)$. 

To explore the role of spin relaxation, we also extend our calculations to the diffusive regime $k_{\eta}/k_{F}=1$ as shown in Fig.~\ref{fig3}. The spiral exchange becomes significantly enhanced, rendering the oscillatory parts into decorative ripples. This can be understood as the decease of the time-reversal symmetry which is broken by the large spin relaxation rate. Note that the RKKY oscillations come from quantum interferences between the patches of the Fermi surfaces related by time-reversal symmetry. Thus, by breaking the symmetry, it is very efficient to weaken the RKKY interaction, as demonstrated by our numerical results. We would like to emphasize that the previous study fails to recognize the important role of the spin relaxation and thus misses out the subtle competition between the spiral exchange and the RKKY interaction completely. 

In conclusion, we demonstrate the important role of spin relaxation in the trilayer magnetic junction and compute the spiral angle between the ferromagnetic layers with different carrier concentrations, (temperature dependent) magnetizations and spin relaxation rates. Our numerical studies show the non-collinear coupling across a DMS thin film is important and will play a crucial role for magnetic junctions at nanoscale. Finally, grant supports from National Science Council in Taiwan are greatly appreciated.


\begin{thebibliography}{00}

\bibitem{Wolf01}
S. A. Wolf, D. D. Awschalom, R. A. Buhrman, J. M. Daughton, S. von Molnar,
M. L. Roukes, A. Y. Chtchelkanova, D. M. Treger
Science {\bf 294}, 1488 (2001).

\bibitem{MacDonald05}
A. H. MacDonald, P. Schiffer and N. Samarth,
Nature Mat. {\bf 4}, 195 (2005).

\bibitem{Akai98}
H. Akai, Phys. Rev. Lett. {\bf 81}, 3002 (1998).

\bibitem{Dietl00}
T. Dietl, H. Ohno, F. Matsukura, J. Cibert and D. Ferrand, Science {\bf 287}, 1019 (2000).

\bibitem{Konig00}
J. K\"onig, H.-H. Lin, and A.~H. MacDonald, Phys. Rev. Lett. {\bf 84}, 5628 (2000).

\bibitem{Schliemann01}
J. Schliemann, J. K\"onig, H.-H. Lin and A.~H. MacDonald, Appl. Phys. Lett. {\bf 78}, 1550 (2001).

\bibitem{Litvinov01}
V.~I. Litvinov and V.~K. Dugaev,
Phys. Rev. Lett. {\bf 86}, 5593 (2001).

\bibitem{Sun04}
S.-J. Sun, S.-S. Chen and H.-H. Lin, 
Appl. Phys. Lett. {\bf 84}, 2862 (2004).

\bibitem{Yang01}
M.-F. Yang, S.-J. Sun and M.-C. Chang,
Phys. Rev. Lett. {\bf 86}, 5636 (2001). 

\bibitem{Konig01}
J. K\"onig, H.-H. Lin and A. H. MacDonald,
Phys. Rev. Lett. {\bf 86}, 5637 (2001).

\bibitem{Sun04a}
S.-J. Sun and H.-H. Lin,
Phys. Lett. A {\bf 327}, 73 (2004).

\bibitem{Datta90}
S. Datta and B. Das, 
Appl. Phys. Lett. {\bf 56}, 665 (1990).



\end{thebibliography}
\end{document}